\documentclass{article}
\usepackage{amssymb,amsmath,cite,bm,eufrak}
\def\e{\EuFrak{e}}

\def\id{\mathbf{1}}
\def\ir{\mathrm{i}}
\def\tr{\mathrm{tr}}
\def\s{\mathrm{s}}
\def\h{\mathrm{h}}
\def\L{\mathrm{L}}

\def\G{\mathcal{G}}
\def\m{\mathrm{m}}
\def\P{\mathrm{P}}

\def\A{\mathcal{A}}

\def\B{\mathcal{B}}
\def\E{\mathcal{E}}
\def\Gr{\mathrm{G}}
\def\Mr{\mathrm{M}}
\def\ri{\mathrm{I}}
\def\rii{\mathrm{I\:\!\!I}}
\begin{document}
\begin{titlepage}
\noindent{\large\textbf{Gauge theory on a space with linear Lie type fuzziness}}

\vspace{\baselineskip}

\begin{center}
Mohammad~Khorrami~{\footnote{mamwad@mailaps.org}}\\
Amir~H.~Fatollahi~{\footnote{fath@alzahra.ac.ir}}\\
Ahmad~Shariati~{\footnote{shariati@mailaps.org}}\\
\textit{ Department of Physics, Alzahra University,
Tehran 1993891167, Iran }
\end{center}

\vspace{\baselineskip}
\begin{abstract}
\noindent The U(1) gauge theory on a space with Lie type noncommutativity is
constructed. The construction is based on the group of translation
in Fourier space, which in contrast to space itself is commutative.
In analogy with lattice gauge theory, the object playing the role of flux of
field strength per plaquette, as well as the action, are constructed.
It is observed that the theory, in comparison with ordinary U(1) gauge theory,
has an extra gauge field component. This phenomena is reminiscent of
similar ones in formulation of SU($N$) gauge theory in space with canonical noncommutativity,
and also appearance of gauge field component in discrete direction of Connes'
construction of the Standard Model.
\end{abstract}

PACS: 11.15.-q, 02.40.Gh

Keyword: Gauge field theories, Noncommutative geometry

\end{titlepage}
\section{Introduction}
In recent years much attention has been paid to the formulation and study
of field theories on noncommutative spaces. The motivation is
partly the natural appearance of noncommutative spaces in some
areas of physics, and the recent one in string theory. In
particular it has been understood that the longitudinal directions
of D-branes in the presence of a constant B-field background
appear to be noncommutative, as seen by the ends of open strings
\cite{9908142,99-2,99-3,99-4}. In this case the coordinates
satisfy the canonical relation
\begin{equation}\label{5.1}
[\hat{x}_\mu,\hat{x}_\nu]=\ir\,\theta_{\mu\,\nu}\,1,
\end{equation}
in which $\theta$ is an antisymmetric constant tensor and $1$
is the unit operator. The theoretical and phenomenological
implications of such noncommutative coordinates have been extensively
studied during the last decade.

One direction to extend studies on noncommutative spaces is to
consider spaces where the commutators of the coordinates are not
constants. Examples of this kind are the noncommutative cylinder
and the $q$-deformed plane (the Manin plane \cite{manin}), the so-called
$\kappa$-Poincar\'{e} algebra \cite{luk} (see also
\cite{majid,ruegg1,ruegg2,amelino0,amelino1,amelino2,kappa,Rob,Gro,chai1,chai2}),
and linear noncommutativity of the Lie algebra type
\cite{snyder} (see also \cite{wess,sasak1,sasak2,sasak3}). In the latter the dimensionless spatial
position operators satisfy the commutation relations of a Lie
algebra:
\begin{equation}\label{5.2}
[\hat{x}_a,\hat{x}_b]= f^c{}_{a\, b}\,\hat{x}_c,
\end{equation}
where $f^c{}_{a\,b}$'s are structure constants of a Lie algebra.
One example of this kind is the algebra SO(3), or SU(2). A special
case of this is the so called fuzzy sphere \cite{madore} (see also 
\cite{presnaj1,presnaj2,presnaj3}), where an irreducible representation of 
the position operators is used which makes the Casimir of the algebra,
$(\hat{x}_1)^2+(\hat{x}_2)^2+(\hat{x}_3)^2$, a multiple of the
identity operator (a constant, hence the name sphere). One can
consider the square root of this Casimir as the radius of the
fuzzy sphere. This is, however, a noncommutative version of a
two-dimensional space (sphere).

In \cite{0612013,fakE1,fakE2} a model was introduced in which the
representation was not restricted to an irreducible one, instead
the whole group was employed. In particular the regular
representation of the group was considered, which contains all
representations. As a consequence in such models one is dealing
with the whole space, rather than a sub-space, like the case of
fuzzy sphere as a 2-dimensional surface. In \cite{0612013} basic
ingredients for calculus on a linear fuzzy space, as well as basic
notions for a field theory on such a space, were introduced. In
\cite{fakE1, fakE2} basic elements for calculating the matrix elements
corresponding to transition between initial and final states,
together with the explicit expressions for tree and one-loop
amplitudes were given. It is observed that models based on
Lie algebra type noncommutativity enjoy three features:
\begin{itemize}
\item They are free from any ultraviolet divergences if the group
is compact.
\item There is no momentum conservation in such
theories.
\item In the transition amplitudes only the so-called
planar graphs contribute.
\end{itemize} The reason for latter is that the non-planar graphs
are proportional to $\delta$-distributions whose dimensions are
less than their analogues coming from the planar sector, and so
their contributions vanish in the infinite-volume limit usually
taken in transition amplitudes \cite{fakE2}. One consequence of
different mass-shell condition of these kinds of theory was explored
in \cite{skf}.

In \cite{kfs} the classical mechanics defined on a space with
SU(2) fuzziness was studied. In particular, the Poisson structure
induced by noncommutativity of SU(2) type was investigated, for
either the Cartesian or Euler parameterization of SU(2) group. The
consequences of SU(2)-symmetry in such spaces on integrability,
was also studied in \cite{kfs}. In \cite{fsk} the quantum mechanics
on a space with SU(2) fuzziness was examined. In particular,
the commutation relations of the position and momentum
operators corresponding to spaces with Lie-algebra
noncommutativity in the configuration space, as well as
the eigen-value problem for the SU(2)-invariant systems were studied.
The consequences of the Lie type noncommutativity of space on thermodynamical
properties have been explored in \cite{shin,Huang,fsmjmp}.

The purpose of this work is to develop a gauge theory on a space
with linear Lie type noncommutativity. In the same line as \cite{0612013,fakE1,fakE2},
the coordinates are taken to be in the regular representation of the Lie algebra.
The gauge theory in space-time with canonical commutation relation (\ref{5.1})
has been the subject of a large number of research attempts; see
\cite{9908142,reviewnc,Sza,Kone} for a list of earlier attempts. One remarkable
feature of the gauge theory on a canonical space is the absence of decoupling
of the U(1) sector of U($N$) from its SU($N$) sector. This phenomenon can be
explained on purely algebraic as well as string theoretic grounds \cite{0006091}.
From the algebraic point of view, the reason can be traced back to the fact that
the SU($N$) sector is simply not closed under the $\star$-product between
group elements, hence starting from pure SU($N$) one always ends up with
extended components for gauge fields, turning SU($N$) to U($N$). This phenomenon
is mentioned here, as it will be seen that a similar phenomenon occurs here as well.
It will be seen that the algebra used to construct the gauge field, requires
an extra component for the gauge filed, here called $H$, even though one is
considering U(1) symmetry. This observation is to some extent reminiscence of
the appearance of gauge field component in the discrete direction of the two-sheet
construction of the Standard Model by A. Connes, based on the Noncommutative
Geometry approach \cite{conn1,conn2,Kast}. In that construction the gauge field component
in the discrete direction between two sheets happens to be the Higgs field in
the Standard Model point of view. As we will see, due to the lattice nature of
space in our model \cite{0612013,fakE1,fakE2}, by analogy with Connes' construction,
the appearance and the role of extra component for the gauge field is quite expected.

The scheme of the rest of this paper is the following. In section 2, a brief
introduction of the group algebra is given, mainly to fix notation. In section 3
the gauge transformation on the so called matter field is introduced and a
recipe is given to construct from a theory which is invariant under
global gauge transformations, one that is invariant under
local gauge transformation. The idea is essentially the same as minimal
coupling applied to lattice theories. In section 4, the connection
on a commutative lattice is reformulated in such a way that can be
generalized to the case of noncommutative spaces. In section 5 the
connection in the noncommutative space is discussed. In section 6
the gauge invariant action corresponding to a (complex) matter field
and a gauge field is presented. Section 7 is devoted to concluding remarks.
\section{The group algebra}
Assume that there exists a unique measure $\mathrm{d} U$ (up to a
multiplicative constant) with the invariance properties
\begin{align}\label{5.3}
\mathrm{d} (V\,U)&=\mathrm{d} U,\nonumber\\
\mathrm{d} (U\,V)&=\mathrm{d} U,\nonumber\\
\mathrm{d} (U^{-1})&=\mathrm{d} U,
\end{align}
for any arbitrary element ($V$) of the group.
For a compact group $G$, such a measure does exist. There are,
however, groups which are not compact but for them as well such
a measure exists. Examples are noncompact Abelian groups.

The meaning of (\ref{5.3}), is that the measure is invariant
under the left-translation, right-translation, and inversion.
This measure, the (left-right-invariant) Haar measure, is unique up to a
normalization constant, which defines the volume of the group:
\begin{equation}\label{5.4}
\int_G\mathrm{d} U=\mathrm{vol}(G).
\end{equation}
Using this measure, one constructs a vector space as follows.
Corresponding to each group element $U$ an element $\e(U)$ is
introduced, and the elements of the vector space are linear
combinations of these elements:
\begin{equation}\label{5.5}
f:=\int\mathrm{d} U\;f(U)\,\e(U),
\end{equation}
The group algebra is this vector space, equipped with the
multiplication
\begin{equation}\label{5.6}
f\bullet g:=\int\mathrm{d} U\,\mathrm{d} V\; f(U)\,g(V)\,\e(U\,V),
\end{equation}
where $(U\,V)$ is the usual product of the group elements. $f(U)$
and $g(U)$ belong to a field (here the field of complex numbers).
It can be seen that if one takes the central extension of the
group U(1)$\times\cdots\times$U(1), the so-called Heisenberg
group, with the algebra (\ref{5.1}), the above definition results
in the well-known star product of two functions, provided $f$ and
$g$ are interpreted as the Fourier transforms of the functions.

So there is a correspondence between functionals defined on the
group, and the group algebra. The definition (\ref{5.6}) can be
rewritten as
\begin{equation}\label{5.7}
(f\bullet g)(W)=\int\mathrm{d} V\;f(W\,V^{-1})\,g(V).
\end{equation}

The delta distribution is defined through
\begin{equation}\label{5.8}
\int\mathrm{d} U\;\delta(U)\,f(U):=f(\id),
\end{equation}
where $\id$ is the identity element of the group.

Next, one can define an inner product on the group algebra.
Defining
\begin{align}\label{5.9}
\langle\e(U),\e(V)\rangle:=&\;\e^\dagger(U)\,\e(V),\nonumber\\
:=&\;\delta(U^{-1}\,V),
\end{align}
and demanding that the inner product be linear with respect to its
second argument and antilinear with respect to its first argument,
one arrives at
\begin{align}\label{5.10}
\langle f,g\rangle&=f^\dagger\,g,\nonumber\\
&=\int\mathrm{d} U\;f^*(U)\,g(U),
\end{align}
where $*$ is complex conjugation.

Finally, one defines a star operation through
\begin{equation}\label{5.11}
f^\star(U):=f^*(U^{-1}).
\end{equation}
This is in fact equivalent to definition of the star operation in
the group algebra as
\begin{equation}\label{5.12}
[\e(U)]^\star:=\e(U^{-1}).
\end{equation}
It is then easy to see that
\begin{align}\label{5.13}
(f\,g)^\star=&g^\star\,f^\star,\\ \label{5.14} \langle f, g\rangle=&
(f^\star\,g)(\id).
\end{align}
One notes that if $f$ is a member of the group algebra, then
$f^\star$ is a member of the group algebra as well, but
$f^\dagger$ is a member of the space dual to the group algebra,
that is, $f^\dagger$ acts linearly on the group algebra (as
a vector space) and results in a scalar (a complex number).
\section{The gauge transformation}
The action for a complex scalar field $\phi$ in a noncommutative space can be
written, similar to \cite{0612013}, as
\begin{equation}\label{5.15}
S_0=\int\mathrm{d}
t\left\{(\dot\phi^\star\bullet\dot\phi)(\id)+[\phi^\star\bullet(O\,\phi)](\id)
-\sum_{j=2}^n\frac{g_j}{2^j\,j!}\,[(\phi^\star\bullet\phi)^j](\id)\right\},
\end{equation}
where
\begin{equation}\label{5.16}
(O\,\phi)(U)=O(U)\,\phi(U),
\end{equation}
and $O(U)$ is a scalar. One also has
\begin{equation}\label{5.17}
O(U^{-1})=O^*(U),
\end{equation}
and that $g_j$'s are real, to make sure that the action is real.
It is seen that the combination $(\phi^\star\bullet\phi)$ is
invariant under the gauge transformation
\begin{equation}\label{5.18}
\phi'=f\bullet\phi,
\end{equation}
provided the function $f$ is unimodular:
\begin{equation}\label{5.19}
f^\star\bullet f=\id.
\end{equation}
To make the whole action gauge invariant, one way is to use
the so called minimal coupling, changing the derivatives into
covariant derivatives. Regarding the time derivative, one uses
$\nabla_0$ instead of $\partial_0$ (the time derivative):
\begin{align}\label{5.20}
\nabla_0\phi&:=\partial_0\phi-\varphi\bullet\phi,\nonumber\\
\nabla_0\phi^\star&:=\partial_0\phi^\star+\phi^\star\bullet\varphi,
\end{align}
where $\varphi$ is pure imaginary:
\begin{equation}\label{5.21}
\varphi^\star=-\varphi,
\end{equation}
and is gauge transformed like
\begin{equation}\label{5.22}
\varphi'=f\bullet\varphi\bullet f^{-1}-f\bullet(\partial_0 f^{-1}).
\end{equation}

Derivatives with respect to the space coordinates are in $O$.
The combination $[\phi^\star\bullet(O\,\phi)]$ is not invariant
under the gauge transformation (\ref{5.18}), as $(O\,\phi')$ is
$[O\,(f\bullet\phi)]$ which is not equal to $[f\bullet(O\,\phi)]$.
This is very much like the case of gauge theories on
commutative spaces, where under gauge transformation (in which
the field is multiplied by a group element valued function)
the derivative of the field is not just multiplied by the
same group element valued function. In the case of gauge theories
on commutative spaces, one way to make the theory gauge invariant
is to substitute differentiation with covariant differentiation,
so that under gauge transformation the covariant derivative of
the field does behave like the field itself, that is,
the covariant derivative is multiplied by the same
group element function. The covariant differentiation operator
is then defined so that under gauge transformation it
undergoes a similarity transformation, with the same group element valued
function. Here a similar thing is done. The operator $O$ is replaced
by an operator $\tilde O$, which is similarity transformed under
the gauge transformation (\ref{5.18}):
\begin{equation}\label{5.23}
\tilde O'\,\phi=f\bullet\tilde O\,(f^{-1}\bullet\phi).
\end{equation}
The operator $O$ contains derivatives with respect to
the space coordinates. So its effect on the field would be
to multiply its Fourier transformation by Fourier variables.
As the Fourier variables corresponding to the space coordinates
are group elements, the effect of $O$ on the field $\phi$ would
be multiplying $\phi(U)$ by some function of $U$. A function of
group can be written as a linear combination of matrix functions
of the group (matrix elements of the irreducible representations of
the group). Hence, defining $\L_\lambda$ through
\begin{equation}\label{5.24}
(\mathrm{L}_\lambda\,\phi)(U):=U_\lambda\,\phi(U),
\end{equation}
where $\lambda$ labels the representations, and $U_\lambda$ is the
$\lambda$ representation of $U$, the action of $O$ would be
\begin{align}\label{5.25}
(O\,\phi)(U)&=\sum_\lambda\tr(a_\lambda\,U_\lambda)\,\phi(U),\nonumber\\
&=:\sum_\lambda\tr[a_\lambda\,(\mathrm{L}_\lambda\,\phi)(U)],
\end{align}
where $a_\lambda$'s are constant matrices.

In the case of gauge theories on commutative spaces, the covariant
derivative is constructed from the ordinary derivative by
introducing another field (the gauge field) so that a combination
of the ordinary derivative and the gauge field is similarity transformed
under the gauge transformation. Here a similar thing is done.
$\tilde O$ is defined based on $O$, using the gauge field $\A$, as
\begin{equation}\label{5.26}
\tilde O\,\phi:=\sum_\lambda\tr[a_\lambda\,\A_\lambda\bullet(\mathrm{L}_\lambda\,\phi)],
\end{equation}
and it would ensure (\ref{5.23}) provided the operators $\A_\lambda$ are
gauge transformed like
\begin{equation}\label{5.27}
\A'_\lambda\bullet(\mathrm{L}_\lambda\,\phi)=f\bullet\A_\lambda\bullet[\mathrm{L}_\lambda\,(f^{-1}\bullet\phi)].
\end{equation}
\section{The connection in the commutative space}
Consider a space $\mathbb{R}^n$. The (space component of the) connection is a vector
with the components $A^a$, which are pure imaginary functions of space (and time), the
gauge transformation of which under the action of a unimodular function $f$ is
\begin{equation}\label{5.28}
\m(A'^a)=\m(f)\,\m(A^a)\,\m(f^{-1})+\m(f)\,\m(\partial^a f^{-1}),
\end{equation}
where $\mathrm{m}(f)$ is an operator which multiplies $f$ by the argument:
\begin{equation}\label{5.29}
[\m(f)]\,g:=f\,g.
\end{equation}
Note that the first term in the above is simply $\m(A^a)$, as the functions
commute with each other.
Alternatively, one can define a covariant derivative $\nabla$ by
\begin{equation}\label{5.30}
\nabla^a:=\partial^a+\m(A^a),
\end{equation}
and write the gauge transformation like
\begin{equation}\label{5.31}
\nabla'^a=\m(f)\,\nabla^a\,\m(f^{-1}).
\end{equation}
The (space-space component of the) curvature is an antisymmetric tensor $B$
of rank 2, related to $A$ through
\begin{equation}\label{5.32}
\m(B^{a\,b})=\m(\partial^a A^b)-\m(\partial^b A^a)+[\m(A^a),\m(A^b)].
\end{equation}
Again the third term vanishes as $A^a$'s are commuting. There is
another way to obtain the curvature, which can be generalized to the
case of noncommutative spaces as well. One considers a set of generators
$T_a$, which can be regarded as the generators of translation in the
direction $a$ of the Fourier space, and build from the connection and
the derivative the operators $\A_\s$ and $\mathrm{L}$ as
\begin{align}\label{5.33}
\A_\s&:=\exp(-\ir\,\ell\,T_a\,A^a),\\ \label{5.34}
\mathrm{L}&:=\exp(-\ir\,\ell\,T_a\,\partial^a),
\end{align}
where $\ell$ is some parameter of the dimension length, and $A^a$'s are
pure imaginary functions. To be more rigorous, one should use
$T_{a\,\lambda}$ instead of $T_a$, where $\lambda$ is a faithful
representation of the Lie algebra corresponding to the group of
translations on the Fourier space. So $\A_\s$ and $\mathrm{L}$ bear the subscript
$\lambda$ as well. One further defines the operator $\P_\s$ as
\begin{equation}\label{5.35}
\P_\s:=\m(\A_\s)\,\mathrm{L}.
\end{equation}
It is seen that if
\begin{equation}\label{5.36}
\P_\s'=\m(f)\,\P_\s\,\m(f^{-1}),
\end{equation}
then for small $\ell$, keeping only the first nonvanishing term
in $\ell$ one arrives at (\ref{5.28}). So one could begin from
$\A_\s$, $\mathrm{L}$, and $\P_\s$, and define $\nabla$ as
\begin{equation}\label{5.37}
T_a\,\nabla^a:=\lim_{\ell\to 0}[(-\ir\,\ell)^{-1}\,(\P_\s-1)].
\end{equation}
Now define $\B_\s$ as
\begin{equation}\label{5.38}
\m(\B_\s):=(\P_\s\otimes\P_\s)\,(\P_\s^{-1}\otimes\P_\s^{-1}),
\end{equation}
in which the tensor product is defined
only on the matrix structure coming from $T_a$'s. As
a consequence, even on ordinary (commutative) spaces,
$\B_\s$ is not identity, as
\begin{align}\label{5.39}
(C\otimes D)^{\alpha\,\beta}{}_{\gamma\,\delta}&=
[(C\otimes 1)\,(1\otimes D)]^{\alpha\,\beta}{}_{\gamma\,\delta},\nonumber\\
&=C^\alpha{}_\gamma\,D^\beta{}_\delta,
\end{align}
while
\begin{equation}\label{5.40}
[(1\otimes D)\,(C\otimes 1)]^{\alpha\,\beta}{}_{\gamma\,\delta}=D^\beta{}_\delta\,C^\alpha{}_\gamma.
\end{equation}
So if the matrix elements of $C$ and $D$ do not commute with each other,
$[(1\otimes D)\,(C\otimes 1)]$ is not necessarily equal to $[(C\otimes 1)\,(1\otimes D)]$,
which is the same as $(C\otimes D)$. This is the case, even if the space is commutative, as
the matrix elements involved contain diferentiation as well as function multiplications, which
do not necessarily commute.

It is also seen from the Baker-Campbell-Hausdorff formula that
the right hand side of (\ref{5.38}) is actually the multiplication
of something, as all of the commutators of derivatives and
function multiplications are function multiplications.

One has
\begin{align}\label{5.41}
\lim_{\ell\to 0}\{(-\ir\,\ell)^{-2}\,[\m(\B_\s)-1]\}=&[T_{a\,\ri}\,\partial^a,T_{b\,\rii}\,\m(A^b)]+
[T_{a\,\ri}\,\m(A^a),T_{b\,\rii}\,\partial^b]\nonumber\\
&+[T_{a\,\ri}\,\m(A^a),T_{b\,\rii}\,\m(A^b)],
\end{align}
where
\begin{align}\label{5.42}
T_{a\,\ri}&:=T_a\otimes 1,\nonumber\\
T_{a\,\rii}&:=1\otimes T_a.
\end{align}
So,
\begin{equation}\label{5.43}
(T_a\otimes T_b)\,\m(B^{a\,b})=\lim_{\ell\to 0}\{(-\ir\,\ell)^{-2}\,[\m(\B_\s)-1]\},
\end{equation}
and again it is seen that the last term in (\ref{5.41}) is vanishing.

Finally, defining
\begin{equation}\label{5.44}
\nabla_0:=\partial_0-\m(\varphi),
\end{equation}
where $\varphi$ is a function (the scalar potential), one can construct $E$
(the space-time component of the curvature) as follows.
First define $\E_\s$ as
\begin{equation}\label{5.45}
\m(\E_\s):=\P_\s\,\nabla_0\,\P_\s^{-1}-\nabla_0.
\end{equation}
It is seen that
\begin{equation}\label{5.46}
\lim_{\ell\to 0}[(-\ir\,\ell)^{-1}\,\m(\E_\s)]=-T_a\,\m(\partial^a\varphi+\partial_0 A^a),
\end{equation}
so that if one defines
\begin{equation}\label{5.47}
E^a:=-(\partial^a\varphi+\partial_0 A^a),
\end{equation}
then
\begin{equation}\label{5.48}
\lim_{\ell\to 0}[(-\ir\,\ell)^{-1}\E_\s]=T_a\,\m(E^a).
\end{equation}

One can extend the definition of $\A_\s$ to $\A$, so that
in the exponent there exists another term which is
a multiple of the scalar matrix:
\begin{equation}\label{5.49}
\A:=\exp(-\ir\,\ell\,T_C\,A^C),
\end{equation}
where capital letters as index can take the value $\h$ in addition to
the values corresponding to the space directions. $T_\h$ is a pure
imaginary multiple of the unit matrix, and $A^\h$ (denoted by $H$)
is a pure imaginary function, being transformed as
\begin{equation}\label{5.50}
H'=f\,H\,f^{-1}.
\end{equation}
$\A$ is a unitary function, but unlike $\A_\s$ is not necessarily unimodular.

Extending the above definition to other quantities (removing
the subscript $\s$), one arrives at
\begin{align}\label{5.51}
\P&:=\m(\A)\,\mathrm{L},\\ \label{5.52}
\m(\B)&:=(\P\otimes\P)\,(\P^{-1}\otimes\P^{-1}),\\ \label{5.53}
\m(\E)&:=\P\,\nabla_0\,\P^{-1}-\nabla_0,
\end{align}
so that,
\begin{align}\label{5.54}
\lim_{\ell\to 0}[(-\ir\,\ell)^{-1}\,(\P-1)]&=T_a\,\nabla^a+T_\h\,\m(H),\\ \label{5.55}
\lim_{\ell\to 0}\{(-\ir\,\ell)^{-2}\,[\m(\B)- 1]\}&=(T_a\otimes T_b)\,\m(B^{a\,b})\nonumber\\
&\quad+(T_a\otimes T_\h-T_\h\otimes T_a)\,\m(\partial^a H),\\ \label{5.56}
\lim_{\ell\to 0}[(-\ir\,\ell)^{-1}\,\m(\E)]&=T_a\,\m(E^a)-T_\h\,\m(\partial_0 H).
\end{align}
Of course these can be written like
\begin{align}\label{5.57}
\lim_{\ell\to 0}[(-\ir\,\ell)^{-1}\,(\P-1)]&=T_C\,\nabla^C,\\ \label{5.58}
\lim_{\ell\to 0}\{(-\ir\,\ell)^{-2}\,[\m(\B)- 1]\}&=(T_C\otimes T_D)\,\m(B^{C\,D}),\\ \label{5.59}
\lim_{\ell\to 0}[(-\ir\,\ell)^{-1}\,\m(\E)]&=T_C\,\m(E^C),
\end{align}
where we have set
\begin{equation}\label{5.60}
\partial^\h:=0.
\end{equation}

Using these, one can write the Lagrangian density for the gauge fields
(and the field $H$) as
\begin{equation}\label{5.61}
\mathcal{L}_\Gr=\frac{\hbar}{2\,g^2}\,\lim_{\ell\to 0}[\ell^{-4}\,\tr(\B-1)+\ell^{-2}\,\tr(\E^2)],
\end{equation}
where the generators $T_C$ have been taken so that
\begin{equation}\label{5.62}
\tr(T_C\,T_D)=-\delta_{C\,D},
\end{equation}
and $g$ is a dimensionless coupling constant. In the more explicit form,
\begin{equation}\label{5.63}
\mathcal{L}_\Gr=\frac{\hbar}{2\,g^2}\,\left[\frac{1}{2}\,B^{a\,b}\,B_{a\,b}-E^a\,E_a
+(\partial^a H)\,(\partial_a H)-(\partial_0 H)^2\right].
\end{equation}
The dimensions of the gauge fields $A$ and $H$ are (length)$^{-1}$, those of $B$ and $E$
are (length)$^{-2}$, so the dimension of the Lagrangian density above is the
dimension of action times (length)$^{-4}$.
The Lagrangian is the Lagrangian density integrated over space, or the value of the
Fourier transformed Lagrangian density calculated at the origin. It is seen that in this
Lagrangian density, the field $H$ does appear but it is not coupled to the
other components of the gauge field.

It is instructive to mention that the action (\ref{5.61}), with the
limit $\ell\to 0$ removed and also without the $H$ component, does in fact
coincide with the continuous time version of the Wilson U(1) lattice gauge theory
\cite{wilson}. Of course in this case, the functions involved are periodic in
momenta, and the integration region of the momenta is compact.

For the so called matter part of the Lagrangian density, one has
\begin{equation}\label{5.64}
\delta_{a\,b}\partial^a\,\partial^b=-\lim_{\ell\to 0}[(-\ir\,\ell)^{-2}\,\tr(\mathrm{L}+\mathrm{L}^{-1}-2)].
\end{equation}
Substituting $\mathrm{L}$ in the right-hand side with its covariant form $\P$,
one arrives at
\begin{align}\label{5.65}
\delta_{C\,D}\,\nabla^C\,\nabla^D&=-\lim_{\ell\to 0} \{(-\ir\,\ell)^{-2}\,\tr[\m(\A)\,\mathrm{L}+\mathrm{L}^{-1}\,\m(\A^{-1})-2]\},\nonumber\\
&=\delta_{a\,b}\,\nabla^a\,\nabla^b+\m(H^2).
\end{align}
So the matter part of the Lagrangian density would be
\begin{align}\label{5.66}
\mathcal{L}_\Mr&=(\dot\phi^*+\phi^*\,\varphi)\,(\dot\phi-\varphi\,\phi)
+\phi^*\,(\delta_{a\,b}\,\nabla^a\,\nabla^b)\phi-\mu^2\,\phi^*\,\phi
-\sum_{j=2}^n\frac{g_j}{2^j\,j!}\,(\phi^*\,\phi)^j\nonumber\\
&\quad+H^2\,\phi^*\,\phi.
\end{align}
This is the usual gauge invariant Lagrangian density for a self interacting scalar field,
with an additional term, the interaction of the field $H$ with the scalar field.

\section{The connection in the noncommutative space}
The (space component of the) connection in the representation $\lambda$ is a nonsingular function
$\A_\lambda$ defined on the group with values in $\mathrm{GL}(\mathbb{V}_\lambda)$, where
$\mathbb{V}_\lambda$ is the linear space on which the representation $\lambda$ acts. So,
\begin{equation}\label{5.67}
\A_\lambda=\int\mathrm{d} U\;\A_\lambda(U)\,\e(U),
\end{equation}
where the integration runs over the group with a Haar measure as the integration measure,
and $\e(U)$ is that basis vector in the group algebra $\G$ corresponding to the group $G$,
which corresponds to the element $U$ of the group $G$. By nonsingular it is meant that there exists a
function $\A^{-1}_\lambda$ such that
\begin{align}\label{5.68}
\A_\lambda\bullet\A^{-1}_\lambda&=\A^{-1}_\lambda\bullet\A_\lambda,\nonumber\\
&=\id_\lambda\,\e(\id).
\end{align}
$\A$ is taken to be unitary:
\begin{equation}\label{5.69}
\A^\dagger=\A^{-1}.
\end{equation}

Another operator is $\mathrm{L}_\lambda$ (the left action in the
representation $\lambda$) acting on $\mathbb{V}_\lambda\otimes\G$:
\begin{equation}\label{5.70}
\mathrm{L}_\lambda\,[v\,\e(U)]:=(U_\lambda\,v)\,\e(U).
\end{equation}
To make the notation simpler, where no explicit use of the representation is made, the subscript
$\lambda$ will be omitted.

Defining
\begin{equation}\label{5.71}
[\m(f)]\,g:=f\bullet g,
\end{equation}
for any two members $f$ and $g$ of the group algebra, and
\begin{equation}\label{5.72}
\mathcal{O} (U,V):=\e^\dagger(U)\,\mathcal{O} \,\e(V),
\end{equation}
for a linear operator $\mathcal{O}$ from $\G$ to $\G$, it is seen for an operator $\mathcal{O}$
there exists a member $g$ of $\G$ such that
\begin{equation}\label{5.73}
\mathcal{O}=\m(g),
\end{equation}
if and only if $\mathcal{O}(U,V)$ is a function of $(U\,V^{-1})$, as
\begin{equation}\label{5.74}
\e^\dagger(U)\,\m(g)\,\e(V)=g(U\,V^{-1}).
\end{equation}
An operator $\mathcal{O}$ is called a function multiplication, if there exists
a $g$ in $\G$ such that (\ref{5.73}) is satisfied. It is easy to see that
a linear combination or product of two
function multiplications are function multiplications. In fact
\begin{equation}\label{5.75}
\m(\alpha_1\,g_1+\alpha_2\,g_2)=\alpha_1\,\m(g_1)+\alpha_2\,\m(g_2),
\end{equation}
where $\alpha_i$'s are constants, and
\begin{equation}\label{5.76}
\m(g_1\bullet g_2)=\m(g_1)\,\m(g_2).
\end{equation}
Also if $g$ is nonsingular,
\begin{equation}\label{5.77}
[\m(g)]^{-1}=\m(g^{-1}).
\end{equation}

The parallel transport operator $\P$ is defined as
\begin{equation}\label{5.78}
\P:=\m(\A)\,\mathrm{L}.
\end{equation}
It is seen that
\begin{equation}\label{5.79}
\P_\lambda(U,V)=\A_\lambda(U\,V^{-1})\,V_\lambda.
\end{equation}
The gauge transformation of the parallel transport operator corresponding to a member $f$ of $\G$
is $\P'$ defined as
\begin{equation}\label{5.80}
\P':=\m(f)\,\P\,\m(f^{-1}).
\end{equation}
One has
\begin{align}\label{5.81}
\P'(U,V)&=\int\mathrm{d} W\,\mathrm{d} X\;f(U\,W^{-1})\,\A(W\,X^{-1})\,X\,f^{-1}(X\,V^{-1}),\nonumber\\
&=\int\mathrm{d} W\,\mathrm{d} X\;f[(U\,V^{-1})\,(W\,V^{-1})^{-1}]\nonumber\\
&\quad\times\A[(W\,V^{-1})\,(X\,V^{-1})^{-1}]\,(X\,V^{-1})\,f^{-1}(X\,V^{-1})\,V.
\end{align}
This shows that there exists a function $\A'$ such that
\begin{equation}\label{5.82}
\P'=\m(\A')\,\mathrm{L}.
\end{equation}
So the gauge transformed connection $\A'$ is defined as
\begin{equation}\label{5.83}
\A'(U):=\int\mathrm{d} W\,\mathrm{d} X\;f(U\,W^{-1})\,\A(W\,X^{-1})\,X\,f^{-1}(X).
\end{equation}

One could attempt to amend (\ref{5.69}) with another condition,
something like unimodularity. Then, using a two dimensional representation
of SU(2), one would be left with three (real) degrees of freedom for the
(space part of) the gauge field. The above gauge transformation, however,
does not preserve such a condition. So even for a unimodular representation,
one should include gauge fields corresponding to nonunimodular matrices.
That's the reason the field $H$ was introduced in the previous section.

To define the (space-space component of the) curvature, we prove that there exists
a function $\B_\lambda$ such that
\begin{equation}\label{5.84}
(\P_\lambda\otimes\P_\lambda)\,(\P^{-1}_\lambda\otimes\P^{-1}_\lambda)=\m(\B_\lambda).
\end{equation}

Denoting the left hand side of (\ref{5.84}) by $\mathrm{H}$, one has
\begin{align}\label{5.85}
\mathrm{H}(U,V)&=
\int\mathrm{d} W\,\mathrm{d} X\,\mathrm{d} Y\;\{[\A_\lambda(U\,W^{-1})\,W_\lambda]\otimes
[\A_\lambda(W\,X^{-1})\,X_\lambda]\}\nonumber\\
&\qquad\times\{[X^{-1}_\lambda\,\A^{-1}_\lambda(X\,Y^{-1})]\otimes
[Y^{-1}_\lambda\,\A^{-1}_\lambda(Y\,V^{-1})]\},\nonumber\\
&=\int\mathrm{d} W\,\mathrm{d} X\,\mathrm{d} Y\;[\A_\lambda(U\,W^{-1})\,(W\,X^{-1})_\lambda\,\A^{-1}_\lambda(X\,Y^{-1})]
\nonumber\\
&\qquad\otimes[\A_\lambda(W\,X^{-1})\,(X\,Y^{-1})_\lambda\,\A^{-1}_\lambda(Y\,V^{-1})],
\end{align}
showing that the left hand side depends on only $(U\,V^{-1})$. So $\B_\lambda$ according
to (\ref{5.84}) is well defined. It is easily seen that this curvature undergoes a similarity
transformation under the gauge transformation:
\begin{equation}\label{5.86}
\B'_\lambda=f\bullet\B_\lambda\bullet f^{-1},
\end{equation}
where $\B'$ is the gauge transformed curvature corresponding to a gauge transformation with the
function $f$.

The time component of the connection is denoted be $\varphi$. $\varphi$
is a member of $\G$, using which $\nabla_0$ (the covariant time derivative) is defined as
\begin{equation}\label{5.87}
\nabla_0:=\partial_0-\m(\varphi).
\end{equation}
$\varphi'$ (the gauge transformed $\varphi$) is defined such that the action of
gauge transformation on $\nabla_0$ be a similarity transformation with $\m(f)$:
\begin{equation}\label{5.88}
\nabla'_0=\m(f)\,\nabla_0\,\m(f^{-1}),
\end{equation}
from which one arrives at
\begin{equation}\label{5.89}
\varphi'=f\bullet\varphi\bullet f^{-1}-f\bullet[\partial_0(f^{-1})].
\end{equation}
Finally, the space-time component of the curvature in the representation $\lambda$ is denoted
by $\E_\lambda$, defined through
\begin{equation}\label{5.90}
\m(\E_\lambda):=\P_\lambda\,\nabla_0\,\P_\lambda^{-1}-\nabla_0.
\end{equation}
It has to be shown that the $\m(\E)$ from (\ref{5.90}) is well defined, that is, the
right hand side of (\ref{5.90}) is a function multiplication. Noting that
$\partial_0$ commutes with $\mathrm{L}$, it is seen that
\begin{equation}\label{5.91}
\P_\lambda\,\nabla_0\,\P_\lambda^{-1}-\nabla_0=\m\{\A_\lambda\bullet[\partial_0(\A_\lambda^{-1})]\}
-\P_\lambda\,\m(\varphi)\,\P_\lambda^{-1}+\m(\varphi).
\end{equation}
One also has
\begin{equation}\label{5.92}
\e^\dagger(U)\,\mathrm{L}_\lambda\,\m(\varphi)\,\mathrm{L}_\lambda^{-1}\,\e(V)=(U\,V^{-1})_\lambda\,\varphi(U\,V^{-1}),
\end{equation}
showing that $\mathrm{L}_\lambda\,\m(\varphi)\,\mathrm{L}_\lambda^{-1}$ is a function multiplication, from which
it turns out that the right hand side of (\ref{5.90}) is a function multiplication.
Using (\ref{5.90}), and the transformation properties of $\P$ and $\nabla_0$, it is seen that
the action of gauge transformation on $\E$ is a similarity transformation with $f$:
\begin{equation}\label{5.93}
\E'_\lambda=f\bullet\E_\lambda\bullet f^{-1}.
\end{equation}
\section{The action}
The gauge action is defined similar to the case of commutative space, the
Lagrangian density for which was defined in (\ref{5.61}). First, one notes that
if the function $X$ is similarity transformed under a gauge transformation:
\begin{equation}\label{5.94}
X'=f\bullet X\bullet f^{-1},
\end{equation}
then the value of that function at identity is gauge invariant:
\begin{align}\label{5.95}
X'(\id)&=\int\mathrm{d} U\,\mathrm{d} V\;f(U^{-1})\,X(U\,V^{-1})\,f^{-1}(V),\nonumber\\
&=\int\mathrm{d} U'\,\mathrm{d} V\;f(V^{-1}\,U'^{-1})\,X(U')\,f^{-1}(V),\nonumber\\
&=\int\mathrm{d} U'\;\delta(U')\,X(U'),\nonumber\\
&=X(\id).
\end{align}
Based on this, the pure gauge field sector of the Lagrangian is defined as
\begin{equation}\label{5.96}
L_\Gr=\frac{\hbar}{2\,g^2}\,[\ell^{-4}\,\tr(\B-1)+\ell^{-2}\,\tr(\E^2)](\id).
\end{equation}
Using
\begin{align}\label{5.97}
\B(\id)=\int\mathrm{d} U\,\mathrm{d} V\,\mathrm{d} W\;&[\A(U\,V^{-1})\,V\,W^{-1}\,\A^{-1}(W)]\nonumber\\
\otimes&[\A(U^{-1})\,U\,V^{-1}\,\A^{-1}(V\,W^{-1})],
\end{align}
and
\begin{align}\label{5.98}
\E(U)=&\int\mathrm{d} V\;\A(U\,V^{-1})\,(\partial^0\A^{-1})(V)\nonumber\\
&+\varphi(U)-\int\mathrm{d} V\,\mathrm{d} W\;\A((U\,V^{-1})\,V\,W^{-1}\,\varphi(V\,W^{-1})\,\A^{-1}(W),
\end{align}
one can write the gauge Lagrangian more explicitly as
\begin{align}\label{5.99}
L_\Gr=\frac{\hbar}{2\,g^2\,\ell^4}\,\bigg\{&
\int\mathrm{d} U\,\mathrm{d} V\,\mathrm{d} W\;\tr[\A(U\,V^{-1})\,V\,W^{-1}\,\A^{-1}(W)]\nonumber\\
\times&\tr[\A(U^{-1})\,U\,V^{-1}\,\A^{-1}(V\,W^{-1})]-[\tr(1)]^2\,\delta(\id)\bigg\}\nonumber\\
&+\frac{\hbar}{2\,g^2\,\ell^2}\,\int\mathrm{d} U\,\tr[\E(U)\,\E(U^{-1})].
\end{align}
It may look that the above Lagrangian is singular, due to
the singular term $\delta(\id)$. It is not. One way to see this,
is to write everything in terms of $(\A-1)$ instead of $\A$, so that
the term proportional to $\delta(1)$ is canceled. Alternatively, one
may use the logarithm of $\A$ as the gauge field, to show manifestly
that the above Lagrangian is not singular.

The matter part of the Lagrangian corresponding to (\ref{5.15}) and (\ref{5.66}) is
\begin{align}\label{5.100}
L_\Mr&=[(\dot\phi^\star+\phi^\star\bullet\varphi)\bullet(\dot\phi-\varphi\bullet\phi)](\id)
+\ell^{-2}\,\{\phi^\star\bullet[\tr(\A\,\mathrm{L}+\mathrm{L}^{-1}\A-2)\,\phi]\}(\id)\nonumber\\
&\quad-\mu^2\,(\phi^\star\bullet\phi)(\id)
-\sum_{j=2}^n\frac{g_j}{2^j\,j!}\,[(\phi^\star\bullet\phi)^j](\id).
\end{align}
\section{Concluding remarks}
An alternative construction of gauge theories on commutative spaces
was discussed, which could be generalized to the case of Lie type
noncommutative spaces. This alternative is basically constructed
in terms of parallel transports. Based on that, a gauge theory on
a noncommutative space of Lie type noncommutativity was
constructed, in which the gauge field is basically some representation
of the group the matrix elements of which are functions. Using this,
and the time component of the gauge field, analogues of the electric
and magnetic fields were constructed, from which a gauge action was
built. Apart from differences proportional to the noncommutativity
parameter, it is seen that one has to introduce an extra degree of
freedom in the gauge field. The reason is that even for the simplest
representation of the group SU(2), starting from a three component
(space) gauge field corresponding to the two dimensional representation of
three generators of the SU(2), applying the gauge transformation introduces
a fourth component corresponding to a generator which commutes
with the other three. In other words, there is no analogue
of unimodularity of two dimensional matrices (for gauge fields)
which is preserved under gauge transformation.
\\[\baselineskip]
\textbf{Acknowledgement}:  This work was
supported by the research council of the Alzahra University.

\end{document}